\documentclass[lettersize,journal]{IEEEtran}
\usepackage{amsmath,amsfonts}
\usepackage{algorithmic}
\usepackage{algorithm}
\usepackage{array}
\usepackage[caption=false,font=normalsize,labelfont=sf,textfont=sf]{subfig}
\usepackage{textcomp}
\usepackage{stfloats}
\usepackage{url}
\usepackage{verbatim}
\usepackage{graphicx}
\usepackage{cite}
\usepackage{booktabs}
\usepackage{orcidlink}

\usepackage{enumitem}
\usepackage{multirow}
\usepackage{bbding}
\usepackage{makecell}

\setitemize[1]{itemsep=0pt,partopsep=0pt,parsep=\parskip,topsep=0pt}
\begin{document}

\title{Latent Factor Modeling with Expert Network for Multi-Behavior Recommendation}

\author{

    Mingshi Yan,
    Zhiyong Cheng,
    Yahong Han,
    and
    Meng Wang,~\IEEEmembership{Fellow,~IEEE}
    
  \thanks{Mingshi Yan and Yahong Han are with the College of Intelligence and Computing, Tianjin University, Tianjin, 300354, China (email: neo.ms.yan@gmail.com; yahong@tju.edu.cn).
  
  Zhiyong Cheng and Meng Wang are with the School of Computer Science and Information Engineering, Hefei University of Technology, Hefei, 230002, China (email: jason.zy.cheng@gmail.com; wangmeng@hfut.edu.cn).
  
  Corresponding Author: Yahong Han.}
  
}

\markboth{Journal of \LaTeX\ Class Files,~Vol.~18, No.~9, September~2020}%
{How to Use the IEEEtran \LaTeX \ Templates}




\maketitle

\begin{abstract}


Traditional recommendation methods, which typically focus on modeling a single user behavior (e.g., purchase), often face severe data sparsity issues. Multi-behavior recommendation methods offer a promising solution by leveraging user data from diverse behaviors. However, most existing approaches entangle multiple behavioral factors, learning holistic but imprecise representations that fail to capture specific user intents.
To address this issue, we propose a multi-behavior method by modeling latent factors with an expert network (\textbf{MBLFE}). In our approach, we design a gating expert network, where the expert network models all latent factors within the entire recommendation scenario, with each expert specializing in a specific latent factor. The gating network dynamically selects the optimal combination of experts for each user, enabling a more accurate representation of user preferences. To ensure independence among experts and factor consistency of a particular expert, we incorporate self-supervised learning during the training process. Furthermore, we enrich embeddings with multi-behavior data to provide the expert network with more comprehensive collaborative information for factor extraction. Extensive experiments on three real-world datasets demonstrate that our method significantly outperforms state-of-the-art baselines, validating its effectiveness.

\end{abstract}



\section{Introduction} \label{Introduction}

\IEEEPARstart{R}{ecommender} systems improve user experience by providing personalized recommendations and facilitating information filtering. Their influence spans across e-commerce, social media, entertainment, and more, shaping consumer behavior and everyday lifestyles. Collaborative filtering (CF)~\cite{DGCF, BPR}, the most widely used technique, leverages user behavior data (e.g., ratings, clicks, purchases) to identify interaction patterns and similarities between users and items for tailored recommendations. However, traditional collaborative filtering typically focuses on a single type of behavior, such as purchase, and faces challenges like data sparsity and cold start issues.

To overcome the limitations of traditional collaborative filtering, which focuses on a single type of behavior, researchers have developed multi-behavior recommendation methods~\cite{Disen-CGCN, MBGCN}. These methods consider various behaviors, such as purchase, click, and rating, across different contexts to better capture user preferences and intentions, thus improving the accuracy and personalization of recommendations. In multi-behavior approaches, user behaviors are typically categorized as either target or auxiliary behaviors based on their relevance. The target behavior, such as purchase, is the primary focus of many online recommender systems, while auxiliary behaviors, like click and collect, represent secondary interactions that provide additional insights into user preferences. Current multi-behavior recommendation methods primarily enhance target behavior predictions by leveraging auxiliary behaviors. Different approaches have been developed to achieve this. For example, MBGCN~\cite{MBGCN} learns separate user representations for each behavior and then aggregates them, and MBSSL~\cite{MBSSL} leverages contrastive learning across the levels of the inter- and intra-behavior to refine the user representations. CRGCN~\cite{CRGCN} and MB-CGCN~\cite{MBCGCN} model inter-behavior relationships to refine user representations.

Despite recent advancements, existing multi-behavior recommendation methods often learn holistic user representations from collaborative information, entangling different factors. This may lead to the following issues: 1) The entangled representation may include factors unrelated to a user's core preferences, introducing noise. For instance, if the director is the dominating factor for a user to watch a movie, mixing in information such as actors or genres may impact the recommendation effectiveness. 2) It becomes difficult to determine which specific factor (e.g., director, actor) drives the recommendation, hindering both the capture of true user intent and the interpretability of the recommendation.

\begin{figure}[t]
    \centering
    \includegraphics[width=\linewidth]{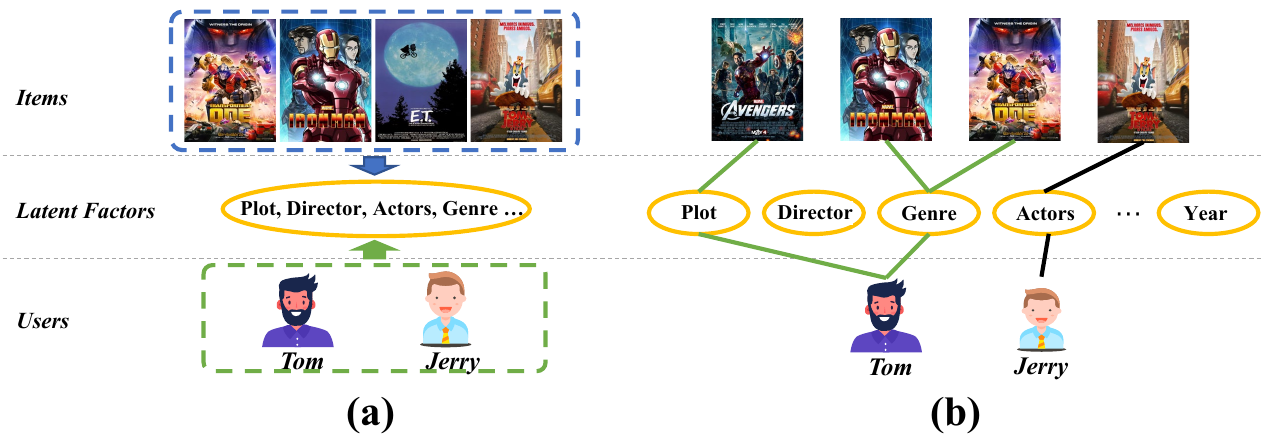}
    \caption{An illustrating of the latent factors.}
    \label{fig:toys}
\end{figure}

To address these challenges, we propose a novel multi-behavior method, MBLFE, extracting latent factors from collaborative information to capture users' true intentions. Latent factors refer to the different underlying intentions or motivations that users exhibit in their interactions~\cite{DGCF, DisenGCN}, and together, they represent the diverse driving forces behind user behaviors. 
Our motivation stems from the observation of shared factors within collaborative information. In the movie recommendation scenario, as illustrated in Fig.~\ref{fig:toys}(a), factors such as plot, director, actors, and genre serve as bridges connecting users and items. These factors reflect both user preferences (e.g., a preference for a specific director or genre) and item attributes (e.g., the director or genre of a movie). Importantly, these factors are not specific to a single user or item but are shared across them. For instance, the "animation" factor may describe multiple movies and align with the preferences of various users, and we refer to these as shared factors. In fact, just as the number of movie genres is finite, the number of shared factors is also limited. Moreover, users typically consider only a few factors during interactions. For example, as shown in Fig.~\ref{fig:toys}(b), Tom watches the anime "\textit{Iron Man}" because he is interested in its plot and genre, while other factors are irrelevant to his preferences. Inspired by this, we model all the latent factors in the entire scenario within the latent space, and then select a limited set of latent factors that determine user interactions as the user representation. This approach enables disentangling, allowing us to more accurately capture user preferences, thereby enhancing both recommendation performance and model interpretability.

Based on the above discussion, we design an expert network to model all the latent factors in the interaction scenario. Each expert focuses on capturing a specific latent factor, and the number of experts matches the number of latent factors in the scenario. This design ensures comprehensive modeling of all latent factors in the scenario, with the number of experts determining the granularity of latent factor modeling. To accurately capture the limited latent factors that drive user interactions, we introduce a gating network that adaptively selects an expert combination for each user, enabling precise capturing of user preferences. Furthermore, the latent factors extracted by the same expert should remain consistent across different users and items. For instance, as shown in Fig.~\ref{fig:toys}(b), Tom prefers the science fiction genre movies, and both "\textit{Iron Man}" and "\textit{Transformers}" belong to the science fiction genre, the latent factor extracted by the specific expert responsible for the movie genre from their representations should be consistent. To ensure such consistency, we adopt a self-supervised approach to align the latent factors extracted by the same expert from different users and items and to maintain the independence of each expert. Good representations are a prerequisite for effective latent factor modeling. To support latent factor learning, we design an embedding enhancement network that leverages the representation learning power of graph convolutional networks (GCNs) to enhance the initialized embeddings. This network mines rich collaborative information from multi-behavior data, providing the expert network with valuable input for factor extraction. Finally, we calculate the correlation between users and items on each latent factor extracted by the experts selected through the gating network. A higher correlation suggests a greater likelihood of user-item interaction on that factor. By integrating these correlations, we can provide users with more precise and interpretable recommendations. Extensive experiments on three real-world datasets demonstrate the effectiveness of our model, achieving significant performance improvements compared to state-of-the-art methods.




In summary, our main contributions are as follows:

\begin{itemize}[leftmargin=*]
\item We propose an expert network to model all latent factors comprehensively. By employing a gating network, our method adaptively selects appropriate experts for users, facilitating precise capture of user interests while filtering out irrelevant information.


\item Before extracting latent factors, we enhance the initialized embeddings using the rich collaborative information from the multi-behavior interactions, ensuring the effectiveness of the gating expert network.

\item Extensive experiments conducted on three real-world datasets demonstrate that our model outperforms state-of-the-art baselines. Additionally, comprehensive ablation studies confirm the effectiveness of our approach. Our code is available on GitHub~\footnote{https://github.com/MingshiYan/MBLFE} for research purpose.

\end{itemize}

The remainder of this paper is organized as follows. Section~\ref{Related Work} introduces the related work and background information. Section~\ref{Methodology} describes the proposed method in detail. Section~\ref{Experiment} provides the experimental setup and results. Finally, Section~\ref{Conclusion} concludes the paper and discusses future directions.






\section{Related Work} \label{Related Work}

\subsection{Multi-behavior Recommendation}



The motivation of multi-behavior recommendation methods is to address the issue of data sparsity~\cite{Disen-CGCN, MBSSL}, which severely hampers the performance of single-behavior recommendation algorithms~\cite{BPR, LightGCN}. Early solutions in this field, such as CMF~\cite{CMF} and BPRH~\cite{BPRH}, utilize matrix factorization techniques, leveraging the sharing of entities between various behaviors during factorization to better manage diverse implicit feedback. In addition, recent approaches incorporate deep learning techniques. These approaches typically employ deep neural networks (DNNs)~\cite{RCF, MATN} or graph neural networks (GNNs)~\cite{GHCF, CKML} to learn user representations from individual behaviors before combining them. A prominent example is MBGCN~\cite{MBGCN}, which applies graph convolutions separately to each behavior and then aggregates the results through weighted combinations to generate the final user representation. Another set of methods aims to enhance user representations by exploring the relationships between different behaviors~\cite{NMTR, ARGO}. For example, CRGCN~\cite{CRGCN} and MB-CGCN~\cite{MBCGCN} establish dependencies between behaviors via sequential modeling, refining user preferences by propagating information across behaviors. Additionally, methods such as MATN~\cite{MATN} and KHGT~\cite{KHGT} incorporate attention mechanisms to encode the interdependencies among behaviors, capturing subtle relationships between various user interactions. Recently, a disentanglement-based multi-behavior method known as Disen-CGCN~\cite{Disen-CGCN} has been proposed. This method focuses on disentangling user multi-behavior representations, aiming to uncover diverse interests across various user behaviors and apply them to the target behavior.

While existing methods have made significant progress, most approaches tend to overlook the diversity of user preferences, often entangling them during the modeling process. It restricts the exploration of users' true intentions. Even though Disen-CGCN incorporates disentanglement, it overlooks the differences in the number of aspects that different users focus on. In this work, we model all potential latent factors and adaptively select the most relevant factors to represent user preferences, thereby enhancing both recommendation performance and interpretability.

\subsection{Disentangled Representation Learning}

Recently, the learning of multi-interests of users has garnered significant attention~\cite{DisMIR, CKML, SimEmb}. Disentangled representation learning~\cite{KMBFD, DCCF, CurCoDis} effectively separates different aspects of user interest from their complex preferences. Existing methods can be categorized into two types: variational autoencoder-based (VAE) method and GNN-based method.

VAE-based methods~\cite{ADDVAE, SEM-MacridVAE, DualVAE} map data into multiple latent spaces. By increasing the weight of the KL divergence term, these methods enforce the latent factors to conform to a prior distribution, compelling the model to identify independent latent factors. For instance, MacridVAE~\cite{MacridVAE} utilizes VAE and KL divergence to achieve macro and micro disentanglement. DualVAE~\cite{DualVAE} combines VAE with attention mechanisms for enhanced representation learning. GNN-based methods~\cite{Disen-GNN, DISGCN, DGCF} extract information at various levels through different graph convolutional layers to disentangle user preferences. For example, DGCF~\cite{DGCF} and DISGCN~\cite{DISGCN} leverage GNNs to learn multiple independent representations before aggregation. DisenGCN~\cite{DisenGCN} infers latent factors of edges in the graph, enabling convolutions on nodes associated with specific factors.

Most studies have focused solely on the independence of different interest aspects without considering the correlations between different users within the same aspect. In MBLFE, we simultaneously ensure the independence and consistency of the experts, allowing for a more accurate modeling of users' multidimensional preferences.
 
\subsection{Mixture of Experts in Recommendation}


The application of Mixture of Experts (MoE) models in recommender systems has become a prominent research focus in recent years~\cite{FAME, M3oE, M3SRec}. The core idea is to provide a flexible and efficient solution for multi-task learning in recommender systems through expert specialization and dynamic routing mechanisms. A representative work in this area is MMoE~\cite{MMoE}, which shares multiple expert networks and designs independent gating networks for each task to dynamically combine expert outputs, thereby modeling inter-task relationships. However, MMoE does not explicitly distinguish between task-specific and shared features, potentially leading to expert redundancy. To address this, PLE~\cite{PLE} introduces a hierarchical routing mechanism, progressively extracting features to mitigate the "seesaw phenomenon" (i.e., performance improvement in some tasks accompanied by degradation in others), demonstrating robustness in industrial-scale recommendation scenarios. Traditional MoE gating networks are typically based on simple linear transformations, which may fail to fully capture complex task relationships. MoME~\cite{MoME} achieves a balance between efficient parameter sharing, flexible task adaptation, and model lightness in multi-task recommender systems by leveraging shared base networks and multi-level masking. MetaHeac~\cite{MetaHeac} incorporates a meta-learning framework, generating expert module parameters via a meta-network, enhancing generalization ability for unseen tasks. In recent years, MoE has been introduced into multi-behavior recommendation methods to enhance multi-behavior modeling. Among these approaches, CIGF~\cite{CIGF} designs a multi-expert network with independent inputs to prevent gradient coupling between different behavior prediction tasks, thereby alleviating the negative transfer problem. COPF~\cite{COPF} adopts a distributed fitting framework with multiple expert networks, leveraging expert coordination in multi-task learning to mitigate negative transfer issues caused by discrepancies in feature and label distributions.

Unlike existing MoE methods, MBLFE focuses on the disentanglement of representations by explicitly modeling shared latent factors through expert networks. Each expert specializes in a specific factor, and contrastive learning ensures independence and consistency among these factors, thereby avoiding information entanglement. This fine-grained factor separation enables more precise capture of users’ core preferences.






\section{Methodology} \label{Methodology}
\subsection{Preliminaries}

In this section, we introduce key concepts, symbols, and the problem settings covered in this work.

Let user $u \in \mathcal{U}$, and item $i \in \mathcal{I}$, with behavior $m \in \mathcal{M}$. For a given behavior $m$, we construct an interaction matrix $\mathcal{R}_m$, where users are represented by rows and items by columns. If a user $u$ interacts with an item $i$ in behavior $m$, then the element in the interaction matrix $\mathcal{R}_m$, denoted as $r_{ui}$, is set to 1; otherwise, $r_{ui}$ is set to 0. User and item embeddings are initialized randomly, with matrices $\boldsymbol{P} \in \mathbb{R}^{d \times |\mathcal{U}|}$ and $\boldsymbol{Q} \in \mathbb{R}^{d \times |\mathcal{I}|}$ representing the initial embeddings for users and items, respectively. The column vectors $\boldsymbol{p_u}$ and $\boldsymbol{q_i}$, derived from $\boldsymbol{P}$ and $\boldsymbol{Q}$ based on user $u$'s and item $i$'s IDs, serve as embeddings for user $u$ and item $i$, respectively.

The problem investigated in this paper can be formally described as follows,

\textbf{Given}: A set of users $\mathcal{U}$, a set of items $\mathcal{I}$, and multi-behavior historical interaction records.

\textbf{Objective}: To learn a prediction function $f$ such that for any user $u \in \mathcal{U}$ and item $i \in \mathcal{I}$, the output $f(u,i)$ represents the probability of user $u$ interacting with item $i$ in the target behavior.

\begin{figure*}[t]
    \centering
    \includegraphics[width=\textwidth]{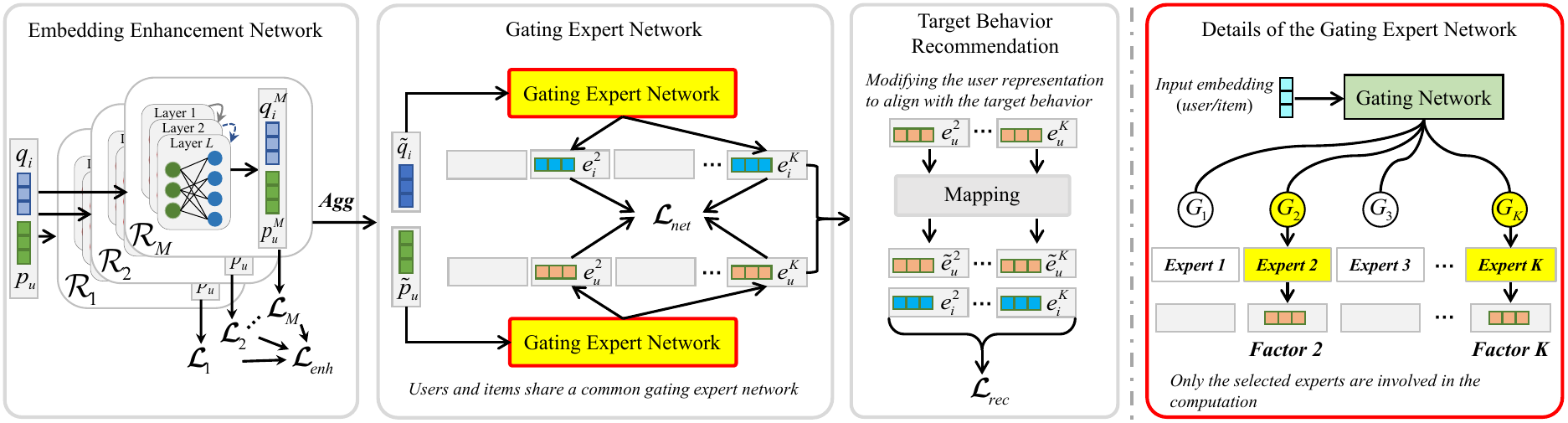}
    \caption{Overview of MBLFE. (The left side shows the overall architecture of MBLFE, while the right side illustrates the internal details of the gating expert network. The initialized embedding is first enhanced by the embedding enhancement network. Then, the gating expert network separately extracts latent factors for both users and items. Finally, the user's latent factors are projected into the target behavior space to generate recommendations. Users and items share a common gating expert network.)}
    \label{fig:global}
\end{figure*}

\subsection{MBLFE Method}




Before delving into the details of MBLFE, let’s revisit the outline of the model. As discussed in Section~\ref{Introduction}, the core of MBLFE lies in modeling all latent factors through the expert network. This approach effectively filters out irrelevant information and captures factor-level correlations between users and items by imposing independence constraints across different experts and consistency constraints within each expert. Furthermore, the adaptive selection of expert combinations for users via a gating network allows for a precise and efficient representation of user preferences. Specifically, we first learn user and item representations from multi-behavior interactions, followed by the extraction of latent factors from these representations. Finally, the factors that reflect user preferences are fine-tuned for target behavior recommendations.

Based on the discussion above, as shown in Fig.~\ref{fig:global}, our method consists of three key components: (1) an \textbf{embedding enhancement network}, which utilizes GCN to mine collaborative information from multi-behavior interactions; (2) a \textbf{gating expert network}, responsible for adaptively extracting latent factors from the learned collaborative information; and (3) a \textbf{target behavior recommendation module}, which projects the extracted factors into the target behavior space to generate recommendations.

In the following sections, we will provide a detailed explanation of the MBLFE method.

\subsubsection{\textbf{Embedding Enhancement Network}}

User and item embeddings are typically initialized randomly (e.g., using Gaussian or uniform distributions)~\cite{SGL, NCL}. The expert network cannot extract latent factors from these noisy embeddings, which lack meaningful information. To address this, we enhance the initialized embeddings before factor extraction. Given the sparsity of target behavior data, we model multi-behavior data to acquire richer collaborative information.

Recently, GCNs have shown great potential for representation learning by incorporating higher-order neighborhood node information~\cite{MRGSRec, MBCGCN}. To simplify our approach, we adopt LightGCN~\cite{LightGCN} to capture collaborative information for each behavior, although more advanced methods, such as NCL~\cite{NCL} or SGL~\cite{SGL}, could also be used as alternatives. 

\textbf{Brief of the LightGCN}. LightGCN employs neighborhood aggregation to update the nodes:
\begin{equation}
  \label{eq:gcn}
  \begin{aligned}
    \boldsymbol p_{u}^{(l+1)} &= \sum_{i \in N_{u}} \frac{1}
    {\sqrt{\left\lvert N_{u} \right\rvert} \sqrt{\left\lvert N_{i} \right\rvert}} \boldsymbol q_{i}^{(l)}, \\ 
    \boldsymbol q_{i}^{(l+1)} &= \sum_{u \in N_{i}} \frac{1}
    {\sqrt{\left\lvert N_{i} \right\rvert} \sqrt{\left\lvert N_{u} \right\rvert}} \boldsymbol p_{u}^{(l)}, 
  \end{aligned}
\end{equation}
where $N_{u}$ and $N_{i}$ represent the first-order neighbors of users and items, respectively. $\boldsymbol p_{u}^{(l+1)}$ and $\boldsymbol q_{i}^{(l+1)}$ denote the embeddings of users and items at the ($l+1$)-th layer of the network, $\boldsymbol p_{u}^{(0)}$ and $\boldsymbol q_{i}^{(0)}$ serve as initial input to the network when $l=0$.

Then aggregate each layer of embeddings for users and items:
\begin{equation}
  \label{eq:norm}
  \begin{aligned}
    \boldsymbol{p}'_{u} = \sum_{l=0}^{L} \boldsymbol{p}_{u}^{(l)}, \quad
    \boldsymbol{q}'_{i} = \sum_{l=0}^{L} \boldsymbol{q}_{i}^{(l)},
  \end{aligned}
\end{equation}
where $L$ denotes the number of GCN layers, $\boldsymbol{p}'_{u}$ and $\boldsymbol{q}'_{i}$ are the outputs of LightGCN, representing the embeddings of users and items, respectively.

\textbf{Embedding Enhancement Learning}. Based on the above LightGCN (Eq. ~\ref{eq:gcn} and ~\ref{eq:norm}), by taking $\boldsymbol{p}_u$ and $\boldsymbol{q}_i$ as the shared input for each behavior, we obtain the sets $\{ \boldsymbol{p}_u^1, \cdots, \boldsymbol{p}_u^m, \cdots, \boldsymbol{p}_u^M\}$ for users and $\{ \boldsymbol{q}_i^1, \cdots, \boldsymbol{q}_i^m, \cdots, \boldsymbol{q}_i^M\}$ for items, where $m$ denotes the $m$-th behavior. 

Similar to LightGCN, we employ the BPR loss function to guide the learning process:
\begin{equation}
  \label{eq:bpr_loss}
  \mathcal{L}_{m} = -\sum_{(u,i,j) \in \mathcal{O}} ln \sigma(\boldsymbol{p}_u^m \boldsymbol{q}_i^m - \boldsymbol{p}_u^m \boldsymbol{q}_j^m),
\end{equation}
where $\mathcal{O}=\{(u,i,j) \in \mathcal{R}_m|r_{ui}=1, r_{uj}=0\}$ represents the observed interactions ($r_{ui} = 1$) and unobserved interactions ($r_{uj} = 0$) in the $m$-th behavior, $\boldsymbol{q}_i^m$ and $\boldsymbol{q}_j^m$ represent the items that have been observed and unobserved in behavior $m$ that interact with user $u$, respectively. The total loss function of the embedding enhancement network is:
\begin{equation}
  \label{eq:enh_loss}
  \mathcal{L}_{enh} = \frac{1}{M} \cdot \sum_{m=1}^{M} \mathcal{L}_m,
\end{equation}
where $M$ represents the total number of behaviors.

Subsequently, we aggregate the embeddings learned from all behaviors to enhance the representations of users and items. Due to variations in user interaction frequencies across different behaviors, there exist discrepancies in their contributions to user representations. Based on users' interaction patterns, we assign corresponding weights to different behaviors. The embedding aggregation operation is as follows:
\begin{equation}
  \label{eq:weight}
  \begin{aligned}
    w_u^m = \frac{n_u^m}{\sum_{j=1}^{M} n_u^j}, \quad
    w_i^m = \frac{n_i^m}{\sum_{j=1}^{M} n_i^j},
  \end{aligned}
\end{equation}
\begin{equation}
  \label{eq:agg}
  \begin{aligned}
    \boldsymbol{\tilde{p}}_{u} = \sum_{m=1}^{M} w_u^m \cdot \boldsymbol{p}_{u}^m, \quad
    \boldsymbol{\tilde{q}}_{i} = \sum_{m=1}^{M} w_i^m \cdot \boldsymbol{q}_{i}^m,
  \end{aligned}
\end{equation}
where $M$ represents the total number of behaviors, $n_u^m$ ($n_i^m$) represents the number of items (users) that user $u$ (item $i$) has interacted with in the behavior $m$, and $w_u^m$ and $w_i^m$ denote the weighting coefficients assigned to the $m$-th behavior for users and items, respectively.

Both $\boldsymbol{\tilde{p}}_{u}$ and $\boldsymbol{\tilde{q}}_{i}$ aggregate collaborative information collected from all behaviors, providing foundational support for modeling in the subsequent expert network. Next, we will explain how latent factors are extracted from them.

\subsubsection{\textbf{Gating Expert Network}}

The gating expert network consists of two components: the \textbf{gating network} and the \textbf{expert network}. The expert network extracts latent factors from input embeddings, with each expert focusing on a specific factor. These latent factors are shared across all users and items, rather than being user-specific or item-specific. For example, movie genres can serve as a factor reflecting both movie characteristics and user interests. The expert network’s role is to extract all potential latent factors from the input embeddings, whether they originate from users or items. The gating network dynamically selects the optimal combination of experts for each user, adapting to their preferences. By analyzing user input embeddings, it identifies the latent factors that most accurately reflect user preferences, capturing their true intent.

Given the embedding $\boldsymbol{\tilde{p}}_{u}$ of user $u$, the process of extracting the latent factor by expert $k$ is as follows:
\begin{equation}
    \label{eq:gate_result}
    \boldsymbol{e}^k_u = G_k(\boldsymbol{\tilde{p}}_{u}) \cdot \boldsymbol{E}_k(\boldsymbol{\tilde{p}}_{u}),
\end{equation}
where $\boldsymbol{e}^k_u$ represents the $k$-th factor of user $u$, $G_k(\cdot)$ and $\boldsymbol{E}_k(\cdot)$ correspond to the $k$-th gate in the gating network and the $k$-th expert in the expert network, respectively.
Note that expert $\boldsymbol{E}_k(\cdot)$ is activated to extract the $k$-th factor only when $G_k(\cdot) > 0$. Next, we will discuss the gating network and the expert network in detail.

\textit{\textbf{Gating Network.}} The gating network dynamically selects experts for computation based on the input, with the number of gates matching the number of experts. Specifically, it assigns a weight to each expert. If an expert's weight exceeds the threshold, the corresponding gate opens, allowing that expert to contribute to the computation. Otherwise, the gate closes. We employ the noisy top-k gating method~\cite{MOE} to implement this mechanism, and introducing noise to ensure that the selected top-k experts adapt more flexibly to the input, thereby improving the model's performance and robustness. The details are as follows:
\begin{equation}
    \label{eq:gate_1}
    \boldsymbol{G}(\boldsymbol{\tilde{p}}_{u}) = T(\boldsymbol{H}(\boldsymbol{\tilde{p}}_{u})), 
\end{equation}
\begin{equation}
    \label{eq:gate_2}
    \boldsymbol{H}(\boldsymbol{\tilde{p}}_{u}) = \boldsymbol{\eta} (\boldsymbol{\tilde{p}}_{u} \cdot \boldsymbol{W}_g + \boldsymbol{S} \cdot \boldsymbol{\delta} (\boldsymbol{\tilde{p}}_{u} \cdot \boldsymbol{W}_{noise})),
\end{equation}
where $\boldsymbol{G}(\cdot)$ represents the gating network, $\boldsymbol{\eta} (\cdot)$ denotes the softmax function, $\boldsymbol{\delta} (\cdot)$ is the softplus function,  $\boldsymbol{W}_g, \boldsymbol{W}_{noise} \in \mathbb{R}^{k \times d}$ are learnable parameters, $\boldsymbol{S}$ represents sampling a random number from a standard normal distribution to introduce noise, and $T(\cdot)$ denotes the top-k gate selector. In real-world scenarios, the number of latent factors that interest different users varies, meaning that the combination of experts selected by each user is different. To address this, we design an adaptive expert selection mechanism. This mechanism dynamically computes a threshold for each user, allowing them to select the most suitable combination of experts based on their individual preferences. Specifically, given a vector $\boldsymbol{x}$ as input, $T(\boldsymbol{x})$ retains only the elements in $\boldsymbol{x}$ that are greater than the mean value of the vector: 
\begin{equation}
    \label{eq:gate_3}
    t_k = \begin{cases}
        x_k,  & \text{if $x_k > \bar{x}$}, \\
              
        0, & \text{otherwise},
    \end{cases}
\end{equation}
where $t_k$ represents the $k$-th element of the output vector $T(\cdot)$ (i.e., $G_k(\cdot)$ in Eq.~\ref{eq:gate_result}), $\bar{x}$ is the mean of all elements in the input vector $\boldsymbol{x}$.

\textit{\textbf{Expert Network.}} For each expert, we apply a two-layer neural network to extract a specific latent factor:
\begin{equation}
    \label{eq:expert}
    \boldsymbol{E}_k(\boldsymbol{\tilde{p}}_{u}) = \boldsymbol{W}_2^k(tanh (\boldsymbol{W}_1^k \boldsymbol{\tilde{p}}_{u} + \boldsymbol{b}_1^k)) + \boldsymbol{b}_2^k,
\end{equation}
where $\boldsymbol{W}_1^k \in \mathbb{R}^{(d/2) \times d}$, $\boldsymbol{W}_2^k \in \mathbb{R}^{d \times (d/2)}$, $\boldsymbol{b}_1^k \in \mathbb{R}^{(d/2)}$ and $\boldsymbol{b}_2^k \in \mathbb{R}^d$ are learnable parameters of expert $k$, $tanh(\cdot)$ denoting the activation function, and $\boldsymbol{E}_k(\cdot)$ means the $k$-th expert of the expert network.

\textit{\textbf{Output.}} According to Eq.~\ref{eq:gate_result}, we obtain the $k$-th latent factor $\boldsymbol{e}^k_u$ of user $u$. Similarly, the $k$-th factor $\boldsymbol{e}^k_i$ of item $i$ can also be obtained. Notably, we can choose only the experts with gate $G_k(\cdot)>0$ to participate in computations, thus reducing the computational load of the model.

For items, instead of applying the gating network for expert selection, we retain all experts to extract latent factors (i.e., fixing $G_k(\cdot)=1$). This approach avoids the risk of having no intersection between the sets of experts selected by both users and items, which would prevent the calculation of correlation between user and item factors, potentially leading to convergence issues. During prediction, item latent factors can be extracted using only the set of experts selected by users, thereby reducing computational costs.

\textit{\textbf{Optimization.}} As described in the introduction, each expert is responsible for extracting a specific latent factor. For instance, expert A and expert B are assigned to extract latent factors of type A and type B, respectively. Therefore, the latent factor extracted by expert A should have a correlation with each other, while maintaining independence between type A and type B latent factors. To ensure consistency within the factors extracted by the same expert and independence between the factors extracted by different experts, we employ contrastive learning for optimization:
\begin{equation}
    \label{eq:infoNCE_user}
    \mathcal{L}_{nce}^{user} = - \sum_{\substack{a, b \in \mathcal{U}\\k \in \mathcal{K}}} log\frac{S(\boldsymbol{e}_a^k,\boldsymbol{e}_b^k)}{S(\boldsymbol{e}_a^k,\boldsymbol{e}_b^k) + \sum \limits_{j \in \mathcal{K} \wedge j \not= k} S(\boldsymbol{e}_a^j, \boldsymbol{e}_b^k)},
\end{equation}
\begin{equation}
    \label{eq:sim_infoNCE_user}
    S(\boldsymbol{e}_a^k, \boldsymbol{e}_b^k) = exp(\frac{{(\boldsymbol{e}_a^k)}^{\top} \boldsymbol{e}_b^k}{\tau}),
\end{equation}
\begin{equation}
    \label{eq:dif_infoNCE_user}
    S(\boldsymbol{e}_a^j, \boldsymbol{e}_b^k) = exp(\frac{{(\boldsymbol{e}_a^j)}^{\top} \boldsymbol{e}_b^k}{\tau}),
\end{equation}
where $\mathcal{K}$ denotes the set of experts, $\mathcal{U}$ represents the user set, and $\tau$ is the temperature hyperparameter. In Eq.~\ref{eq:infoNCE_user}, we enforce independence and consistency constraints by minimizing the correlation between factors extracted by different experts, while simultaneously maximizing the correlation between factors extracted by the same expert. The constraints on items are similar. The total loss of contrastive learning is:
\begin{equation}
    \label{eq:infoNCE}
    \mathcal{L}_{nce} = \mathcal{L}_{nce}^{user} + \mathcal{L}_{nce}^{item}.
\end{equation}

The gating expert network is optimized using self-supervised learning, without needing to specify which expert is chosen or emphasizing the correlation between users and items on specific latent factors. Instead, the focus is entirely on interaction prediction. This approach enables the model to adaptively select the appropriate experts during the learning process, ensuring that several of the most relevant latent factors are identified for each user. Cross-entropy loss is employed for this optimization:
\begin{equation}
    \label{eq:log_loss}
    \mathcal{L}_{log} = -\sum \bar{y} \cdot log(y) + (1-\bar{y}) \cdot log (1-y),
\end{equation}
\begin{equation}
    \label{eq:sum_score}
    y = \sigma (\sum_{k \in \mathcal{S}} ({(\boldsymbol{e}_u^k)}^{\top} \boldsymbol{e}_i^k)),
\end{equation}
where $\bar{y} \in \{0, 1\}$ represents the ground truth, $\bar{y}=1$ when a user interacts with an item, and $\bar{y}=0$ otherwise. $\boldsymbol{e}_u^k$ and $\boldsymbol{e}_i^k$ represent the $k$-th factors of user $ u $ and item $ i $, respectively.
$\sigma(\cdot)$ is the sigmoid function.
$\mathcal{S}$ denotes the set of experts that user $u$ selected.
The objective is to train the gating expert network without needing to consider specific behavior types. To have more data for training, the training set is formed by merging interaction records from all behaviors. 

The total loss of the gating expert network is:
\begin{equation}
    \label{eq:net}
    \mathcal{L}_{net} = \mathcal{L}_{log} + \alpha \cdot \mathcal{L}_{nce},
\end{equation}
where $\alpha \in (0, 1)$ is a hyperparameter.

\subsubsection{\textbf{Target Behavior Recommendation}}
In the previous section, since we utilize interaction data from all behaviors as the training set to optimize the gating expert network, the latent factors of users we learned are not behavior-specific. Therefore, we individually project each user's latent factor into the target behavior space for fine-tuning. The projection of the $k$-th latent factor is:
\begin{equation}
    \label{eq:pfe}
    \boldsymbol{\tilde{e}}_u^k =  (\boldsymbol{e}_u^k)^{\top} \boldsymbol{W}_{map},
\end{equation}
where $\boldsymbol{W}_{map}$ is the projection matrix. We calculate the correlation between users and items for each latent factor extracted by the experts selected by the users. These correlations are then combined to generate personalized recommendations:
\begin{equation}
    \label{eq:pred_score}
    z_{ui} = \sum_{k \in \mathcal{S}} ({(\boldsymbol{\tilde{e}}_u^k)}^{\top} \boldsymbol{e}_i^k),
\end{equation}
where $\mathcal{S}$ denotes the set of experts that user $u$ selected, $\boldsymbol{e}_i^k$ represents the $k$-th factor of item $i$, derived from Eq.~\ref{eq:gate_result}. Notably, we did not apply projection operations analogous to those used for user representations to the item representations. Intuitively, item representations encapsulate their inherent features and attributes, which remain consistent and unchanged across diverse behavioral scenarios. Therefore, rather than projecting the item features, we directly utilize the latent features learned by the gating expert network for recommendation.

We employ the BPR loss function for optimization:
\begin{equation}
  \label{eq:pred_loss}
  \mathcal{L}_{rec} = -\sum_{(u,i,j) \in \mathcal{O}} ln \sigma(z_{ui} - z_{uj}),
\end{equation}
where $\mathcal{O}$ represents the target behavior interaction data, $z_{ui}$ and $z_{uj}$ denote the correlation scores of the user $u$ with the observed item $i$ and the unobserved item $j$ in the target behavior, respectively.

\subsubsection{\textbf{Model Training}}
We optimize the model end-to-end using multi-task learning, where the overall loss function is formulated as:
\begin{equation}
  \label{eq:total_loss}
  \mathcal{L} = \mathcal{L}_{enh} + \mathcal{L}_{net} + \mathcal{L}_{rec} + \gamma \cdot \left\lVert \boldsymbol \Theta \right\rVert_{2},
\end{equation}
where $\gamma$ is a hyperparameter, $\boldsymbol \Theta$ represents all of the parameters in our model, and $\left\lVert \cdot \right\rVert_{2}$ means $L_2$ regularization. 

Please note that, ideally, different weights should be assigned to $\mathcal{L}_{enh}$, $\mathcal{L}_{net}$, and $\mathcal{L}_{rec}$ to balance the optimization process. However, doing so would increase the complexity of hyperparameter tuning, which is not the focus of this work. Therefore, we omit it.

\subsection{\textbf{Model Analysis}}

\subsubsection{\textbf{Time Complexity Analysis}} \label{analys}

The time complexity of our method is mainly determined by three components: the embedding enhancement network, the gating expert network, and the target behavior recommendation. Specifically, the embedding enhancement, implemented by GCNs, has a time complexity of $O(B(LE + Nd))$, where $B$ represents the number of behaviors, $L$ is the number of GCN layers, $E$ denotes the number of edges, $N$ means the number of nodes, and $d$ is the embedding size. The gating expert network, which includes the gating network and expert network, has a time complexity of $O(N(K + 1)d^2)$, where $K$ denotes the number of experts. The target behavior recommendation contributes a time complexity of $O(Nd^2)$. Thus, the total time complexity of our method is $O(B(LE + Nd) + N(K + 2)d^2)$. Notably, since the gating network controls the selection of experts for each user, only several chosen experts participate in the computations, keeping the overall computational complexity acceptable.


\subsubsection{\textbf{Discussion}}

The work related to our method is the disentanglement approach. Existing methods typically represent user preferences across different aspects by learning multiple independent embeddings. However, these approaches neglect the correlations among users within the same aspect and fail to account for differences in the number of aspects that interest different users. In contrast, our method models all latent factors and adaptively selects a few of the most relevant factors to represent each user. By sharing latent factors across users and ensuring consistency within the same factors, our method effectively captures commonalities at the latent factor level, leading to significant improvements in both recommendation accuracy and interpretability. 

Moreover, our method can link specific latent factors with keywords in retrieval tasks, seamlessly integrating recommendation and retrieval. This approach enables more precise personalized recommendations while enhancing the controllability and interpretability of the results. Consequently, our method not only offers significant advantages in personalized services but also holds broad potential for various applications.

\section{Experiment} \label{Experiment}

\begin{table}[t]
  \caption{Statistics of four datasets (the bold text represents the target behavior).}
  \label{tab:dataset}
  \resizebox{\columnwidth}{!}{
    \begin{tabular}{cccccc}
      \toprule
      \textbf{Dataset} & \textbf{\# User} & \textbf{\# Item} & \textbf{\# Interation} & \textbf{Behavior Type} \\
      \midrule
      \textbf{Tmall}  & 41,738 & 11,953 & $2.3\times10^6$   & \{\textit{click, collect, cart, \textbf{purchase}}\} \\
      \textbf{Taobao} & 48,749 & 39,493 & $2.0\times10^6$   & \{\textit{click, cart, \textbf{purchase}}\} \\
      \textbf{Yelp}   & 19,800 & 22,734 & $1.4\times10^6$   & \{\textit{tips, dislike, neutral, \textbf{like}}\} \\
      \bottomrule
    \end{tabular}
  }
\end{table}

\subsection{Experiment Settings}

\subsubsection{\textbf{Dataset}}

\begin{table*}[htb]
  \caption{Overall performance across three datasets ("\textit{Impr.}" represents the relative performance improvement compared to the best baseline, the underline represents the suboptimal performance, and "$\boldsymbol{\ast}$" denotes statistically significant improvements, as determined by two-tailed t-tests with $p<0.05$).}
  \label{tab:overall}
  \resizebox{\textwidth}{!}{
	\begin{tabular}{crcccccccccccccr}
	\toprule
	\multirow{2}{*}{\textbf{Dataset}} & \multirow{2}{*}{\textbf{Metric}} & \multicolumn{3}{c}{\textbf{Single-behavior}}       & \multicolumn{10}{c}{\textbf{Multi-behavior}}    & \multirow{2}{*}{\textit{\textbf{Impr.}}} \\ \cmidrule(lr){3-5} \cmidrule(lr){6-15}
	                      &         & \textbf{MF} & \textbf{LightGCN} & \textbf{DGCF} & \textbf{MBGCN} & \textbf{S-MBRec} & \textbf{CRGCN} & \textbf{CIGF} & \textbf{MB-CGCN} & \textbf{PKEF} & \textbf{Disen-CGCN} & \textbf{COPF} & \textbf{BCIPM} & \textbf{MBLFE}  & \\ \hline  
	\multirow{4}{*}{\textbf{Tmall}}   & \textbf{HR@10}   & 0.0230   & 0.0393   & 0.0396   & 0.0549   & 0.0694   & 0.0840   & 0.0859  & 0.1073   & 0.1118  & 0.1217  & 0.1229  & \underline{0.1414} & $\boldsymbol{0.1452^*}$ & 2.69\%  \\ 
					                  & \textbf{NDCG@10} & 0.0124   & 0.0209   & 0.0208   & 0.0285   & 0.0362   & 0.0442   & 0.0426  & 0.0416   & 0.0630  & 0.0581  & 0.0683  & \underline{0.0741}  & $\boldsymbol{0.0787^*}$ & 6.21\%  \\ 
									  & \textbf{HR@20}   & 0.0316   & 0.0538   & 0.0535   & 0.0799   & 0.1009   & 0.1238   & 0.1247  & 0.1296   & 0.1770  & 0.1766  & 0.1718  & \underline{0.2049} & $\boldsymbol{0.2061^*}$ & 0.59\%  \\ 
					                  & \textbf{NDCG@20} & 0.0144   & 0.0243   & 0.0241   & 0.0345   & 0.0438   & 0.0540   & 0.0496  & 0.0537   & 0.0552  & 0.0679  & 0.0803  & \underline{0.0897} & $\boldsymbol{0.0932^*}$ & 3.90\%  \\ \hline

	\multirow{4}{*}{\textbf{Taobao}}  & \textbf{HR@10}   & 0.0178   & 0.0254   & 0.0244   & 0.0434   & 0.0571   & 0.1152   & 0.0671  & 0.0989   & 0.1097  & 0.1101  & 0.1043  & \underline{0.1292} & $\boldsymbol{0.1343^*}$ & 3.95\%  \\ 
					                  & \textbf{NDCG@10} & 0.0101   & 0.0138   & 0.0134   & 0.0259   & 0.0331   & 0.0629   & 0.0494  & 0.0470   & 0.0627  & 0.0644  & 0.0585  & \underline{0.0716} & $\boldsymbol{0.0795^*}$ & 11.03\%  \\ 
									  & \textbf{HR@20}   & 0.0286   & 0.0374   & 0.0353   & 0.0803   & 0.0890   & 0.1468   & 0.0897  & 0.1379   & 0.1219  & 0.1421  & 0.1515  & \underline{0.1834} & $\boldsymbol{0.1898^*}$ & 3.49\%  \\ 
					                  & \textbf{NDCG@20} & 0.0133   & 0.0169   & 0.0162   & 0.0355   & 0.0401   & 0.0772   & 0.0551  & 0.0602   & 0.0632  & 0.0731  & 0.0704  & \underline{0.0853} & $\boldsymbol{0.0912^*}$ & 6.92\%  \\ \hline

	\multirow{4}{*}{\textbf{Yelp}}    & \textbf{HR@10}   & 0.0327   & 0.0400   & 0.0347   & 0.0356   & 0.0353   & 0.0367   & 0.0501  & 0.0355   & 0.0423  & 0.0399  & 0.0398  & \underline{0.0502} & $\boldsymbol{0.0514^*}$ &  2.39\%  \\ 
					                  & \textbf{NDCG@10} & 0.0159   & 0.0202   & 0.0178   & 0.0183   & 0.0173   & 0.0178   & 0.0240  & 0.0164   & 0.0210  & 0.0197  & 0.0191  & \underline{0.0244} & $\boldsymbol{0.0255^*}$  &  4.51\%  \\
									  & \textbf{HR@20}   & 0.0498   & 0.0577   & 0.0582   & 0.0557   & 0.0518   & 0.0590   & 0.0835  & 0.0577   & 0.0764  & 0.0595  & 0.0661  & \underline{0.0874} & $\boldsymbol{0.0893^*}$ & 2.17\%  \\ 
					                  & \textbf{NDCG@20} & 0.0203   & 0.0246   & 0.0237   & 0.0252   & 0.0217   & 0.0211   & 0.0323  & 0.0215   & 0.0302  & 0.0223  & 0.0257  & \underline{0.0337} & $\boldsymbol{0.0365^*}$ & 8.31\%  \\
	\bottomrule

	\end{tabular}
  }
   
\end{table*}

Our model is evaluated on three datasets:
\begin{itemize}[leftmargin=*]

\item \textbf{Tmall}\footnote{https://tianchi.aliyun.com/dataset/140281}: This dataset originates from Tmall, a comprehensive online shopping platform, and includes four types of behaviors: \textit{click}, \textit{collect}, \textit{cart}, and \textit{purchase}. \textit{Purchase} is the target behavior, with the other three as auxiliary behaviors.

\item \textbf{Taobao}\footnote{https://tianchi.aliyun.com/dataset/649}: Derived from Taobao, another online shopping platform, this dataset comprises three behavior types: \textit{click}, \textit{cart}, and \textit{purchase}. \textit{Purchase} is the target behavior, while \textit{click} and \textit{cart} are auxiliary behaviors.

\item \textbf{Yelp}\footnote{https://www.kaggle.com/yelp-dataset/yelp-dataset}: Collected from Yelp, a U.S.-based local online search platform, this dataset features user \textit{tips} and \textit{ratings}. As per prior studies~\cite{GNMR, MATN}, behaviors are categorized based on \textit{ratings} into \textit{dislike}, \textit{neutral}, and \textit{like}, with \textit{like} as the target behavior and the other two as auxiliary behaviors.

\end{itemize}

For all datasets, we follow previous works~\cite{NMTR, MBGCN, MBCGCN} to retain the original interactions and remove duplicates. Detailed statistical information about the datasets is provided in Tab.~\ref{tab:dataset}.

\subsubsection{\textbf{Evaluation}}
We utilize two widely adopted evaluation metrics, HR@k and NDCG@k, to assess model performance. During the evaluation, we follow a full-ranking protocol~\cite{DiffRec, CRGCN, MBCGCN}, which ranks all items not interacted with by a user. Additionally, we employ the commonly used leave-one-out validation method~\cite{NMTR, MBGCN} for evaluation.

\subsubsection{\textbf{Parameter Settings}}
During training, to ensure fairness, we uniformly set the embedding size to 64. The Adam~\cite{Adam} optimizer is applied for gradient descent, and hyperparameters are tuned via grid search. Specifically, we set the number of GCN layers to 2, the learning rate is selected from $\{1e^{-2}, 3e^{-3}, 1e^{-3}, 1e^{-4} \}$, the temperature parameter $\tau$ from \{0.1, 0.3, 0.5, 0.7, 0.9\}, the total number of experts $K$ from the range 4 to 32, the parameter $\alpha$ from $\{1e^{-1}, 1e^{-2}, 1e^{-3} \}$, and the regularization coefficient $\gamma$ from $\{1e^{-2}, 1e^{-3}, 1e^{-4} \}$. The key parameters in the baseline methods are carefully optimized according to the guidelines of the literature.

\subsubsection{\textbf{Baselines}}
To evaluate the model, we compare MBLFE with three representative single-behavior methods and nine multi-behavior methods.

\textbf{Single-behavior methods:}

\begin{itemize}[leftmargin=*]
\item \textbf{MF}~\cite{BPR}. This is a matrix factorization-based collaborative filtering method, which learns collaborative information in the latent space.
\item \textbf{LightGCN}~\cite{LightGCN}. This method simplifies GCN operations while retaining the neighborhood aggregation component, enhancing the performance of GCN.
\item \textbf{DGCF}~\cite{DGCF}. DGCF is a graph-based disentanglement method that employs graph convolution to learn multiple independent aspects of user preferences.
\end{itemize}

\textbf{Multi-behavior methods:}

\begin{itemize}[leftmargin=*]
\item \textbf{MBGCN}~\cite{MBGCN}. This method employs GCN to model each behavior individually before merging them. It also models item-item graphs to tackle cold-start issues.
\item \textbf{S-MBRec}~\cite{SMBREC}. This method introduces a star-style contrastive learning framework aimed at identifying the commonalities between auxiliary behaviors and the target behavior.
\item \textbf{CRGCN}~\cite{CRGCN}. This method models multiple behaviors sequentially, continuously refining user preferences through message propagation.
\item \textbf{CIGF}~\cite{CIGF}. This paper proposes a multi-behavior recommendation framework based on a compressed interaction graph, which integrates a gated expert network to facilitate knowledge sharing.
\item \textbf{MB-CGCN}~\cite{MBCGCN}. This method improves upon CRGCN by removing the residual connections and simplifying the learning process.
\item \textbf{PKEF}~\cite{PKEF}. The method introduces a projection approach to separate noise in auxiliary behaviors, addressing the issue of negative transfer.
\item \textbf{Disen-CGCN}~\cite{Disen-CGCN}. This method extends MB-CGCN by incorporating disentanglement, separately disentangling the representations of users within each behavior.
\item \textbf{COPF}~\cite{COPF}. The method designs a distributed fitting multi-expert network to coordinate multi-task learning, alleviating the negative transfer problem caused by differences in feature and label distributions.
\item \textbf{BCIPM}~\cite{BCIPM}. The method introduces a behavior-contextualized item preference network to capture crucial information about user-item interactions within specific behavioral contexts.
\end{itemize}

\subsection{Overall Performance}
We first present the results of our method compared to baseline methods on three datasets in Tab.~\ref{tab:overall}. From the statistical results, we have the following observations: 



\textit{Superiority of Multi-Behavior Methods}: The experimental results indicate that the multi-behavioral methods generally perform better than single-behavior methods. By leveraging auxiliary behaviors, multi-behavioral approaches significantly improve the learning of user preferences and effectively address challenges like data sparsity, which are common in single-behavior methods.

\textit{Effectiveness of MBLFE}: Our method outperforms the baseline models across all metrics, demonstrating the effectiveness of its design. MBLFE models all latent factors and adaptively selects relevant factors as user preferences. Each expert specializes in learning a specific latent factor, filtering out irrelevant information. By imposing independence constraints across different factors and consistency constraints within the same factor, the model uncovers correlations between users and items at the factor level. In contrast, existing multi-behavior methods primarily explore collaborative information without decoupling latent factors. Although Disen-CGCN introduces disentanglement, it only learns a few different aspects of user preferences and does not explore the relationships of users and items on these aspects. Among the three methods of incorporating expert networks, MBLFE, CIGF, and COPF, MBLFE utilizes expert networks to achieve disentanglement, enabling more precise separation of diverse dimensions in user interests. CIGF focuses on enhancing knowledge sharing efficiency, though its high-order graph interactions may introduce noise or interest aliasing issues. COPF relies on behavioral constraints and task coordination to optimize holistic representations, but this approach may implicitly retain the influence of noise.

\textit{Single-Behavior Method Analysis}: Among single-behavior methods, LightGCN performs the best due to the strong representation learning capabilities of GCN, which enhances user representations by utilizing the graph aggregation properties of GCN to explore higher-order neighbor information. DGCF models users' diverse interests and has also shown good performance, validating the effectiveness of disentangled representation learning.

\textit{Multi-Behavior Method Analysis}: Among multi-behavior methods, MBGCN and S-MBRec adopt a parallel modeling approach, focusing solely on modeling the data. They learn user preferences from each behavior independently and then aggregate them. CRGCN, MB-CGCN, PKEF, and Disen-CGCN employ a sequential modeling approach, which considers the relationships between different behaviors, leading to better performance. Disen-CGCN stands out by disentangling different aspects of user preferences, further enhancing performance and highlighting the importance of decoupling users' diverse interests. Additionally, BCIPM focuses on identifying key factors in user interactions while avoiding noise from auxiliary behaviors, confirming that entangled representations can introduce significant noise from auxiliary behavior.

\subsection{Ablation Study}
To further validate the effectiveness of MBLFE, we conduct the following ablation experiments.




\textbf{(1) Evaluate embedding enhancement network}:
\begin{itemize}[leftmargin=*]
\item \textbf{\textit{w/o EEN}}:  The embedding enhancement network is removed, and randomly initialized embeddings are used as input for the gating expert network.
\item \textbf{\textit{w/o $\mathcal{L}_{enh}$}}: The loss function $\mathcal{L}_{enh}$ is removed.
\end{itemize}

\textbf{(2) Evaluate expert network}:
\begin{itemize}[leftmargin=*]
\item \textbf{\textit{w/o EXP}}: Remove the gating expert network and directly project the enhanced embeddings into the target behavior space for recommendations.
\item \textbf{\textit{w/o $\mathcal{L}_{nce}$}}: Remove the loss function $\mathcal{L}_{nce}$.
\item \textbf{\textit{w/o $\mathcal{L}_{log}$}}: Remove the loss function $\mathcal{L}_{log}$.
\end{itemize}

\textbf{(3) Evaluate gating network}:
\begin{itemize}[leftmargin=*]
\item \textbf{\textit{w/o GN}}: Remove the gating network, allowing both users and items to select all experts.
\item \textbf{\textit{w/ all GN}}: Both users and items utilize gating network to select the appropriate experts.
\end{itemize}

\begin{table}[t]
    \caption{Results of the ablation experiment (all metrics are evaluated at \textit{top@10}, "\textit{w/o}" means without and "\textit{w/}" means with).}
    \label{tab:abl}
    \resizebox{\linewidth}{!}{
    \begin{tabular}{lcccccc}
    \toprule
    \multirow{2}{*}{\textbf{Method}} & \multicolumn{2}{c}{\textbf{Tmall}} & \multicolumn{2}{c}{\textbf{Taobao}} & \multicolumn{2}{c}{\textbf{Yelp}}  \\ \cmidrule(lr){2-3} \cmidrule(lr){4-5} \cmidrule(lr){6-7}
        & \multicolumn{1}{c}{\textbf{HR}} & \multicolumn{1}{c}{\textbf{NDCG}} & \multicolumn{1}{c}{\textbf{HR}} & \multicolumn{1}{c}{\textbf{NDCG}} & \multicolumn{1}{c}{\textbf{HR}} & \multicolumn{1}{c}{\textbf{NDCG}} \\ \hline
    \textbf{\textit{w/o EEN}}                 & 0.0116                    & 0.0056                      & 0.0071                    & 0.0037                      & 0.0073                    & 0.0036                      \\
    \textbf{\textit{w/o $\mathcal{L}_{enh}$}} & 0.1306                    & 0.0702                      & 0.1133                    & 0.0663                      & 0.0418                    & 0.0199                      \\ \hline
    \textbf{\textit{w/o EXP}}                 & 0.0745                    & 0.0376                      & 0.0512                    & 0.0276                      & 0.0423                    & 0.0210                      \\  
    \textbf{\textit{w/o $\mathcal{L}_{log}$}}                 & 0.1227                    & 0.0666                      & 0.0825                    & 0.0490                      & 0.0313                    & 0.0151                      \\ 
    \textbf{\textit{w/o $\mathcal{L}_{nce}$}}                 & 0.1418                    & 0.0759                      & 0.1201                    & 0.0659                      & 0.0470                    & 0.0224                      \\ \hline
    \textbf{\textit{w/ all GN}}               & 0.0890                    & 0.0463                      & 0.0718                    & 0.0400                      & 0.0249                    & 0.0121                      \\
    \textbf{\textit{w/o GN}}                  & 0.1341                    & 0.0730                      & 0.1224                    & 0.0723                      & 0.0480                    & 0.0236                      \\ \hline
    \textit{\textbf{MBLFE}}                    & \textbf{0.1452}           & \textbf{0.0787}             & \textbf{0.1343}           & \textbf{0.0795}             & \textbf{0.0514}           & \textbf{0.0255}             \\
    \bottomrule
    \end{tabular}
    }
\end{table}

The experimental results are shown in Tab.~\ref{tab:abl}. Overall, the results demonstrate that MBLFE outperforms other models, validating the effectiveness of each module's design.

\textbf{Effectiveness of the embedding enhancement network}: After removing the embedding enhancement network (\textit{w/o EEN}), performance significantly declined across all three datasets. This is attributed to the lack of meaningful information in the randomly initialized embeddings, which impairs the learning of the gated expert network, underscoring the importance of embedding enhancement. The superior performance of \textit{w/o $\mathcal{L}_{enh}$} over \textit{w/o EEN} further supports this. Despite the absence of explicit supervision in \textit{w/o $\mathcal{L}_{enh}$}, GCNs can still extract collaborative information for the gating expert network learning. Comparing \textit{w/o $\mathcal{L}_{enh}$} and \textit{MBLFE} shows that supervised learning of embedding enhancement yields further improvements, indicating that more accurate collaborative information benefits the extraction of latent factors.

\textbf{Effectiveness of the expert network}:  A significant performance decline is observed when the gating expert network is removed, as demonstrated by the comparison between \textit{w/o EXP} and \textit{MBLFE}, highlighting the gating expert network's effectiveness. Without this network, the model lost the ability to extract users' true intentions from entangled representations, leading to suboptimal performance. After removing loss functions $\mathcal{L}_{nce}$ (\textit{w/o $\mathcal{L}_{nce}$}) and $\mathcal{L}_{log}$ (\textit{w/o $\mathcal{L}_{log}$}), a noticeable decrease in performance occurs, validating the effectiveness of these constraints. Without $\mathcal{L}_{nce}$, the model loses the independence and consistency constraints for the experts, reducing its expressive capacity. Similarly, without $\mathcal{L}_{log}$, the model cannot effectively learn the optimal expert combinations, also degrading performance. 

\textbf{Effectiveness of the gating network}: Comparing to \textit{MBLFE}, both \textit{w/ all GN} and \textit{w/o GN} demonstrate inferior performance. The reduced performance of \textit{w/ all GN}, in particular, is due to the application of the gating network to both users and items, which occasionally leads to an empty intersection of the expert sets selected for users and items. It hinders the model's ability to converge effectively. The distinction between \textit{w/o GN} and \textit{MBLFE} lies in \textit{w/o GN}'s removal of the gating network. In \textit{w/o GN}, retaining all experts causes the final user representations to include numerous factors with low or irrelevant relevance to users. These factors introduce noise, leading to the observed performance degradation.

\subsection{Gating Expert Network Study}

The gating expert network is pivotal in MBLFE, as it enables the model to adaptively select relevant latent factors to represent user preferences. To further validate its effectiveness, we conduct additional experimental experiments.

\subsubsection{\textbf{Expert Selection Analysis}}
\begin{figure}[ht]
    \centering
    \includegraphics[width=\linewidth]{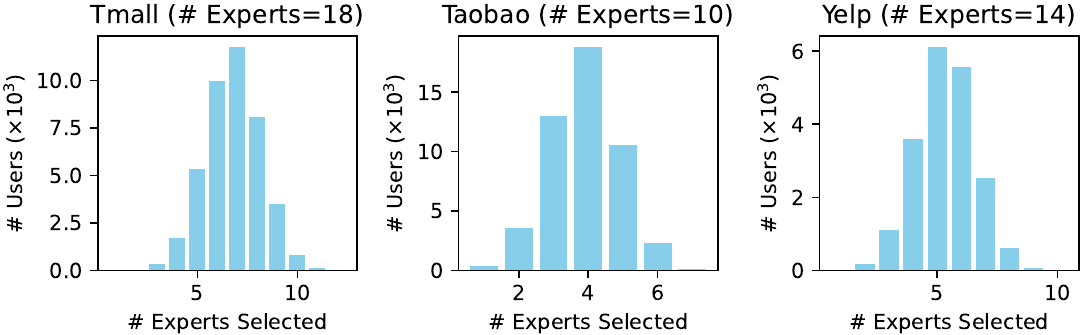}
    \caption{Statistics on the number of experts selected by users (the $x$-$axis$ and $y$-$axis$ represent the number of experts and the number of users, respectively, and "\# Experts=18" means that the total number of experts is 18).}
    \label{fig:expert}
\end{figure}

The gating network controls expert selection for users. Fig.~\ref{fig:expert} provides a statistical analysis of the number of experts chosen across different datasets.

From the statistical results, we observe that the number of experts selected by users follows a normal distribution across the three datasets. Specifically, in the Tmall dataset, most users selected 7 experts, while in the Taobao and Yelp datasets, users chose 4 and 5 experts, respectively, significantly less than half of the total number of available experts. This observation validates the effectiveness of our top-k selection strategy. In real-world scenarios, users who concentrate on either an exceptionally small or large number of factors constitute a minority, whereas the majority clusters around the center of the distribution. By adaptively selecting the top-k experts, our model better mirrors this real-world pattern, enabling it to provide more accurate and tailored expert combinations for each user, thereby improving the effectiveness of personalized services.

\subsubsection{\textbf{Visualization of Latent Factors Distribution}}
\begin{figure}[ht]
    \centering
    \includegraphics[width=\linewidth]{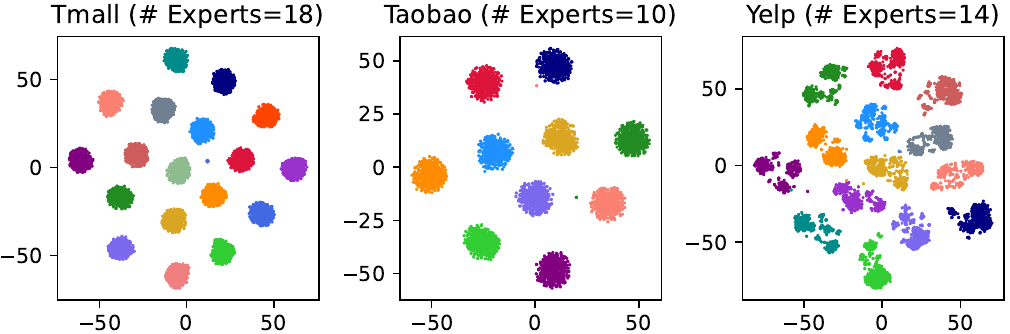}
    \caption{Visualization of latent factors distribution extracted by different experts (each color corresponds to the factor extracted by a specific expert, and "\# Experts=18" means that the total number of experts is 18).}
    \label{fig:scatter}
\end{figure}

Each expert is responsible for extracting a specific factor, making it crucial to ensure both independence and consistency among these factors. To evaluate this, we visualize the distribution of all factors. In detail, we randomly sample 500 user latent factors from the outputs of each expert, perform dimensionality reduction using t-SNE, and project the results into a two-dimensional space for visualization. In the resulting figure, each color represents the outputs associated with a specific expert. The results are shown in Fig.~\ref{fig:scatter}.

From the visualization results, we observe that the preference factors across the three datasets exhibit clear clustering, with the number of clusters corresponding to the total number of experts. This suggests that the factors extracted by different experts remain independent, while those extracted by the same expert display high consistency. This outcome can be attributed to the independence constraints applied to the preference factors through contrastive learning. These constraints ensure that user preference factors within the same cluster align well with item features, leading to more accurate recommendations and enhanced interpretability.

\subsection{Total Expert Number Analysis}
\begin{figure}[ht]
    \centering
    \includegraphics[width=\linewidth]{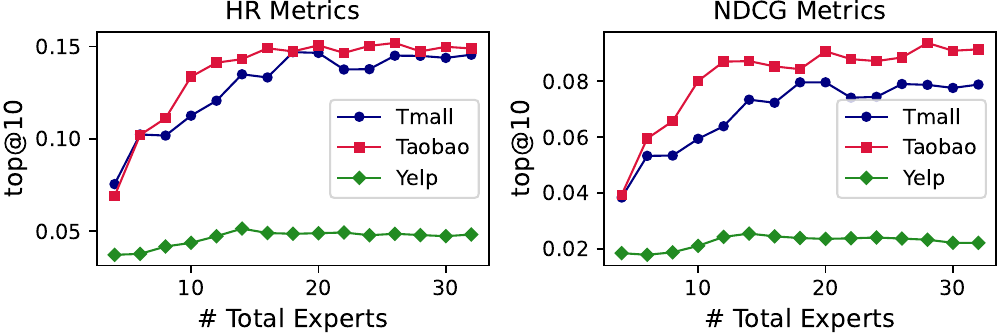}
    \caption{Impact of the total number of experts on model performance.}
    \label{fig:total_experts}
\end{figure}

The total number of experts in the gating expert network determines the granularity of the latent factors. A higher number of experts allows for more fine-grained factors. To analyze the impact of varying expert numbers, Fig.~\ref{fig:total_experts} illustrates how the model performance changes as the number of experts increases.


In Fig.~\ref{fig:total_experts}, as the total number of experts increases, both the HR@10 and NDCG@10 metrics demonstrate significant improvement until a certain threshold is reached. This trend can be attributed to the finer granularity of latent factors achieved with a higher number of experts, resulting in more accurate recommendations. However, since the number of latent factors is limited, exceeding this critical threshold can lead to redundancy. Furthermore, as the total number of experts rises, computational costs also increase. Therefore, it is crucial to strike a balance between recommendation performance and computational costs in practical applications, emphasizing the importance of selecting the optimal total number of experts.

In the experiments conducted in this paper, considering various factors, the number of experts in the Tmall, Taobao, and Yelp datasets was set to 18, 10, and 14, respectively, to achieve optimal recommendation performance.

\subsection{Efficiency Analysis}
In addition to the accuracy of the recommendations, operational efficiency is a key metric for evaluating the performance of the model. In Section~\ref{analys}, we analyze the complexity of the model in time, assessing its efficiency in terms of computational resource consumption. To provide a comprehensive evaluation of the model's operational efficiency, this section presents the average training times per epoch and compares them with four representative methods, accompanied by a time-based statistic to further assess the model's efficiency. The experimental environment is configured as follows: \textbf{CPU}: Intel(R) Xeon(R) CPU E5-2650 v4@2.20GHz, \textbf{GPU}: GeForce RTX 2080 Ti Rev. A, \textbf{Batch size}: 1,024, \textbf{Embedding size}: 64.

\begin{table}[htb]
  \caption{Training time statistics per epoch (the unit is in "seconds").}
  \label{tab:time}
  \resizebox{\columnwidth}{!}{
    \begin{tabular}{rccccc}
    \toprule
    \textbf{Dataset}                  & \textbf{LightGCN} & \textbf{CRGCN} & \textbf{MBGCN} & \textbf{BCIPM} & \textbf{MBLFE} \\ \hline
    \textbf{Tmall}                    & 3.58      & 17.33          & 106.72    & 579.49         & 116.03          \\
    \textbf{Taobao}                   & 4.31      & 17.69          & 111.15    & 588.94         & 133.72          \\
    \textbf{Yelp}                     & 2.33      & 8.42           & 42.79     & 479.49         & 59.33           \\
    \bottomrule

    \end{tabular}
  }
\end{table}

Tab.~\ref{tab:time} compares the training time per epoch of MBLFE with LightGCN, CRGCN, MBGCN, and BCIPM across three datasets. LightGCN is a lightweight single-behavior method, CRGCN and MBGCN are two GCN-based multi-behavior methods, and BCIPM is the baseline that performs best in recommendation accuracy. The results show that, during training, MBLFE’s average time consumption is between that of CRGCN and BCIPM, and is similar to that of MBGCN. The core of the MBLFE model involves multiple expert networks and the extraction of latent features from multi-behavior data, which inherently requires more computational resources compared to simpler models like LightGCN or CRGCN. However, it is important to emphasize that MBLFE utilizes a gating network mechanism to activate specific expert networks, minimizing unnecessary computational overhead, which results in performance comparable to MBGCN.

Moreover, the time complexity of MBLFE is considered acceptable due to its capability to provide significantly improved recommendation accuracy and interpretability. The trade-off between slightly increased training time and enhanced performance in terms of accuracy and feature disentanglement is well-justified, especially in practical applications where model performance and user satisfaction are paramount.

\section{Conclusion} \label{Conclusion}
This paper introduces an expert network that models all latent factors and employs a gating network to adaptively select optimal expert combinations for users, enabling the extraction of relevant latent factors. Additionally, we impose independence and consistency constraints on the expert network, allowing the model to establish connections between users and items at a granular level of latent factors, thereby improving both recommendation accuracy and interpretability. Experiments conducted on three datasets validate the effectiveness of our proposed model. For future work, we plan to explore more effective gating mechanisms to select appropriate expert combinations for users and investigate more efficient and controllable disentanglement methods to further enhance recommendation performance and interpretability.
\section{Acknowledgements}
  This research is supported by the National Natural Science Foundation of China under Grant 62376186 and 62272254.

\ifCLASSOPTIONcaptionsoff
  \newpage
\fi
\bibliographystyle{IEEEtran}
\bibliography{reference}

\begin{IEEEbiography}
[{\includegraphics[width=1in,height=1.25in,clip,keepaspectratio]{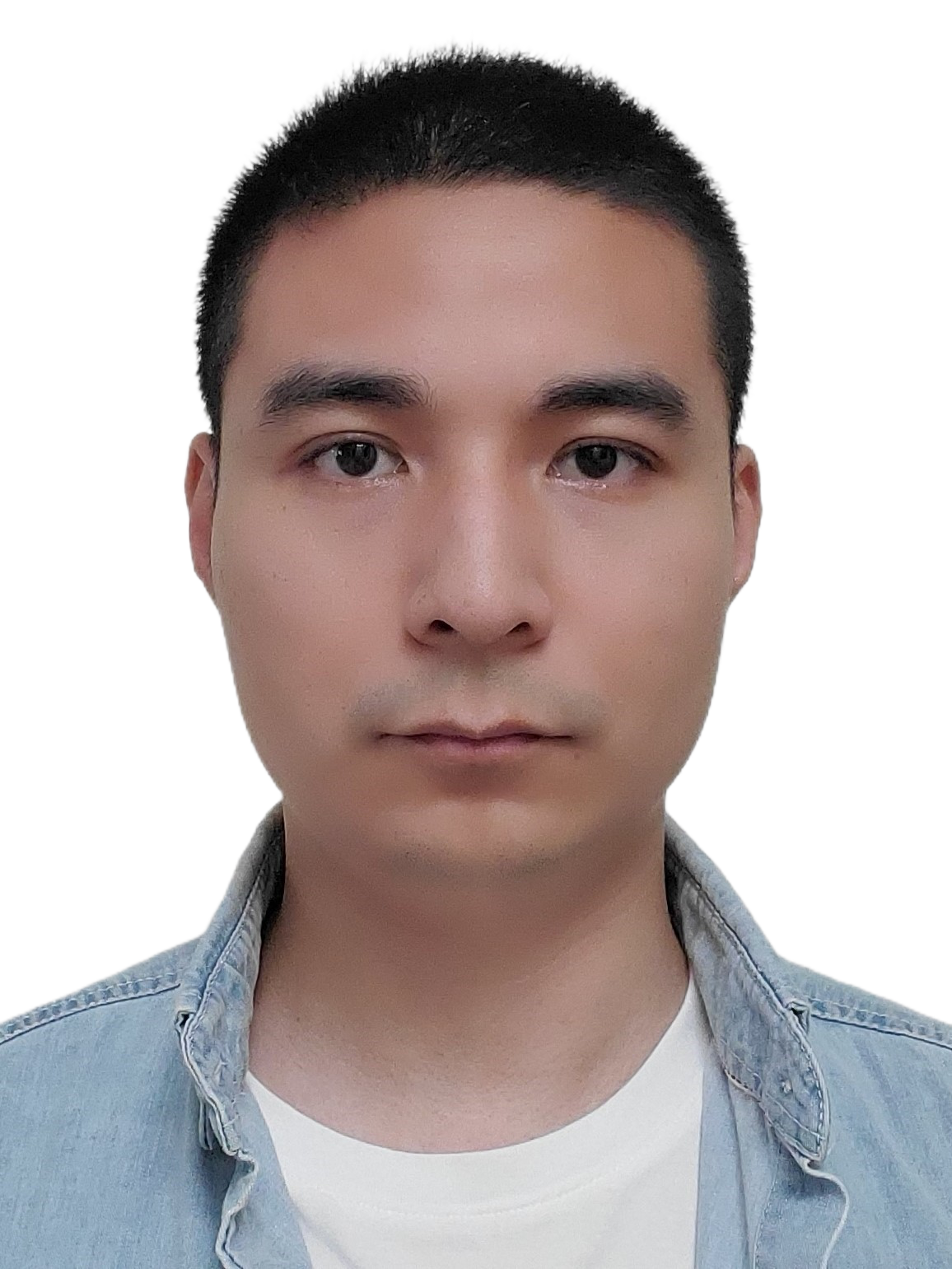}}]{Mingshi Yan}
currently pursuing the Ph.D. degree majoring in computer science and technology with the College of Intelligence and Computing, Tianjin University, Tianjin, China. His main research interests include artificial intelligence and personalized information recommendation.
\end{IEEEbiography}

\begin{IEEEbiography}
[{\includegraphics[width=1in,height=1.25in,clip,keepaspectratio]{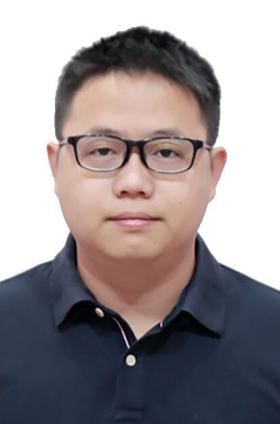}}]{Zhiyong Cheng}
received the Ph.D. degree in computer science from Singapore Management University, Singapore, in 2016. He is currently a professor at the School of Computer Science and Information Engineering, Hefei University of Technology. He has long been engaged in research on multimedia information retrieval and recommendation systems. He has led and participated in over ten projects, including National Natural Science Foundation of China projects and key projects of Shandong Province. He serves as an associate editor for several international journals, including IEEE TCSS, and has served as an AC/SPC for major international conferences such as ACM MM and SIGIR. He has received the "Wu Wenjun Artificial Intelligence Excellent Young Scholar Award" and has been nominated for Best Paper Awards at SIGIR 2019 and ACM MM 2019.
\end{IEEEbiography}

\begin{IEEEbiography}
[{\includegraphics[width=1in,height=1.25in,clip,keepaspectratio]{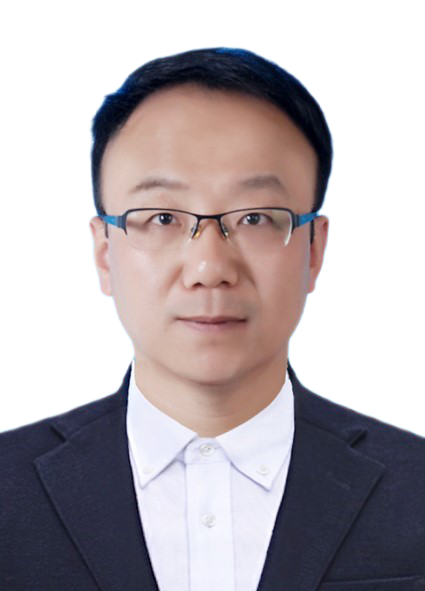}}]{Yahong Han}
received the Ph.D. degree from Zhejiang University, Hangzhou, China, in 2012. He is currently a Professor with the College of Intelligence and Computing, Tianjin University, Tianjin, China. From November 2014 to
November 2015, he visited Prof. Bin Yu’s Group, UC Berkeley, as a Visiting Scholar. His current research interests include multimedia analysis, computer vision, and machine learning.
\end{IEEEbiography}

\begin{IEEEbiography}
[{\includegraphics[width=1in,height=1.25in,clip,keepaspectratio]{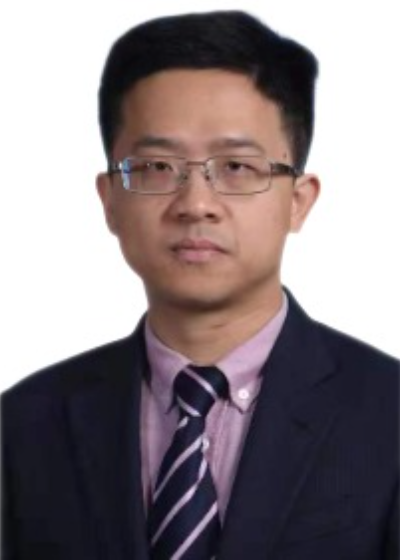}}]{Meng Wang}
(Fellow, IEEE) received the BE and PhD degrees from the Special Class for the Gifted Young and the Department of Electronic Engineering and Information Science, University of Science and Technology of China (USTC), Hefei, China, respectively. He is a professor with the Hefei University of Technology, China. His current research interests include multimedia content analysis, search, mining, recommendation, and large-scale computing. He received the best paper awards successively from the 17th and 18th ACM International Conference on Multimedia, the best paper award from the 16th International Multimedia Modeling Conference, the best paper award from the 4th International Conference on Internet Multimedia Computing and Service, and the best demo award from the 20th ACM International Conference on Multimedia.
\end{IEEEbiography}

\end{document}